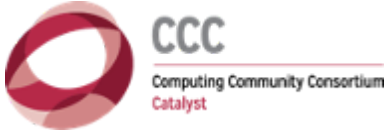

# Computing Research Challenges in Next Generation Wireless Networking

*A Computing Community Consortium (CCC) Quadrennial Paper*

*Elisa Bertino (Purdue University), Daniel Bliss (Arizona State University), Daniel Lopresti (Lehigh University), Larry Peterson (Princeton University), and Henning Schulzrinne (Columbia University)*

## Introduction

By all measures, wireless networking has seen explosive growth over the past decade. Fourth Generation Long Term Evolution (4G LTE) cellular technology has increased the bandwidth available for smartphones, in essence, delivering broadband speeds to mobile devices. The most recent 5G technology is further enhancing the transmission speeds and cell capacity, as well as, reducing latency through the use of different radio technologies and is expected to provide Internet connections that are an order of magnitude faster than 4G LTE. Technology continues to advance rapidly, however, and the next generation, 6G, is already being envisioned. 6G will make possible a wide range of powerful, new applications including holographic telepresence, telehealth, remote education, ubiquitous robotics and autonomous vehicles, smart cities and communities (IoT), and advanced manufacturing (Industry 4.0, sometimes referred to as the Fourth Industrial Revolution), to name but a few [1, 2, 3, 4]. The advances we will see begin at the hardware level and extend all the way to the top of the software "stack."

Artificial Intelligence (AI) will also start playing a greater role in the development and management of wireless networking infrastructure by becoming embedded in applications throughout all levels of the network. The resulting benefits to society will be enormous.

At the same time these exciting new wireless capabilities are appearing rapidly on the horizon, a broad range of research challenges loom ahead. These stem from the ever-increasing complexity of the hardware and software systems, along with the need to provide infrastructure that is robust and secure while simultaneously protecting the privacy of users. Here we outline some of those challenges and provide recommendations for the research that needs to be done to address them.

### Security and Privacy - a Quick Look at 5G

Securing 5G is already a challenging and complex task. The 5G network protocol stack consists of multiple layers, e.g., the physical layer, radio resource control (RRC) layer, and non-access stratum (NAS) layer. Each layer in turn has its own protocols to implement its procedures, such as the protocols for connecting and disconnecting devices to and from the network and for paging devices to deliver notifications of incoming calls and text messages. As a result, along with the vulnerabilities in the inherited functionalities from 3G and 4G networks, the new technologies and protocols added in 5G introduce new attack surfaces that have not yet been fully analyzed with respect to security and user privacy. Having a robust 5G ecosystem will require designing protocols that are able to achieve their promised security and privacy guarantees even in the presence of adversaries, and include the ability to reason about stateful protocols that employ cryptographic constructs (e.g., [5]).

While systematic analysis helps in identifying the root causes of vulnerabilities, it is also vital to design efficient mitigation techniques and secure solutions to protect the next-generation cellular networks against advanced threats. These are defenses against fake base stations, fake emergency alerts, identity exposure attacks, and certain forms of side-channel attacks. However, deploying security techniques for wide-spread use in cellular networks is also challenging due to the nature of the cellular network ecosystem and the divergent incentives of its stakeholders.

### Security, Privacy and Reliability for 6G - A Formidable Research Challenge

The initial visions for 6G share key ideas: (a) use of virtualization technologies across all layers of communication systems; (b) pervasive deployment of AI techniques (e.g., to incorporate semantics [1]); (c) tight integration of communication, computation, caching, and control (C4); and (d) as a way of coping with ever-increasing complexity, deep programmability with a focus on high-level intents and verifiability [2].

Fielding 6G will be extremely challenging due to tight interconnections among its different components, increasing demands for reliability due to its inclusion in critical applications, and extensive deployment of AI, which has its own security and privacy issues. Initial research directions with respect to security, privacy, and reliability include [4]: incorporating security at the physical layer, such as the use of quantum key distribution (QKD) schemes for future optical wireless communication, and authentication by a physical layer signature (such as RF fingerprinting), among others. Taking a broader view, security of 6G can be seen as security of a complex system of systems.

6G networking with a built-in security, privacy, and reliability lifecycle will require not only adopting existing security practices but also developing new ones, including the ability to identify anomalies and vulnerabilities, and then to contain attacks and failures in order to minimize disruptions that would have catastrophic consequences. AI will play an important role in addressing such challenges, but specialized hardware, component redundancy, and diversity will also be critical for successful and safe implementation.

**Further into the Future**

The development of 6G and beyond will enable a wider range of capabilities. Because the flexibility and efficiency of hardware are quickly improving, the traditional constraints on communications waveforms and protocols will become less rigid. Currently, power-efficient implementations of radio processing are typically done in single-purpose full-custom application-specific integrated circuits (ASICs). This type of processing is at least one hundred times more efficient than the processor in a standard computer. However, such ASICs are also inflexible. A new class of application-domain-specific processors are being developed that approach ASIC efficiency, but are relatively flexible. By employing these more flexible processors, a revolution in how spectrum is used becomes available by enabling powerful and efficient software-defined radios (SDRs). As an example of resulting potential advances, the traditional barriers between types of users of RF and mmWave signals can be removed. The idea of RF convergence or spectrum sharing becomes viable with systems jointly using the same RF energy and spectrum to employ functions such as communications, radar, positioning, navigation, and timing (PNT), for example. These joint technologies have broad and significant potential for a variety of applications including improving the safety of autonomous ground and air vehicles, capabilities of human-machine interfaces, and benefits of personal health monitoring.

**Research Opportunities in Deep Programmable Networks**

Traditionally, wireless networks had fixed functions, largely defined by standards and vendors. Future networks will become programmable by their owners and users to suit their needs [7]. This offers a way to manage the complexity of future networks. Network owners will be able to specify the desired behavior at the "top," and the specification will be partitioned and compiled "down" to dictate how the network is controlled and packets are processed. If network owners and users can tailor the network to suit their needs, they likely will, leading to novel designs that are more reliable, secure, and performant.

Deep programmability—and the opportunity to build networks that run autonomously under verifiable, closed-loop control—is already happening, with data-center fabrics and the backbones interconnecting those data centers at the vanguard. But the biggest opportunity, both in terms of technical challenges and societal impact, is the access network. This is a particularly critical moment in time for the mobile cellular network, with the early stage transition to 5G opening the door to deep programmability being truly end-to-end.

Making this huge transition for existing networks would be challenging enough, but 5G and beyond has laid out an ambitious roadmap for the next decade, including support for (i) a massive Internet of Things, including devices with ultra-low energy, ultra-low complexity, and ultra-high density; (ii) mission-critical control, including ultra-low latency, ultra-high predictability, and extreme mobility; and (iii) enhanced mobile broadband, with extreme capacity and data rates. Disaggregation opens the door to deep programmability, but using that opportunity to bring verifiable closed-loop control to demanding applications on complex networks requires a significant research investment [2].

**Recommendations**

Next generation wireless networking ecosystems, including 5G, and the envisioned 6G and beyond, are complex and involve a large number of parties with differing interests and perspectives. The challenges span all the way from the lowest hardware layer to the top of the software stack. These networks offer tremendous potential to society, but research investments are urgently needed in security, privacy, reliability, deployability and programmability if we are to reap the benefits. Beyond the work taking place in academia, this will also require collaborating with industry across multiple sectors, including network providers, equipment manufacturers, software and service companies, as well as regulatory and standardization bodies.

5GReasoner: A Property-Directed Security and Privacy Analysis Framework for 5G Cellular Network Protocol. ACM Conference on Computer and Communications Security 2019: 669-684

[7] Nate Foster, et. al. Using Deep Programmability to Put Network Operators in Control. *ACM SIGCOMM Computer Communication Review*, October 2020.

*This white paper is part of a series of papers compiled every four years by the CCC Council and members of the computing research community to inform policymakers, community members and the public on important research opportunities in areas of national priority. The topics chosen represent areas of pressing national need spanning various subdisciplines of the computing research field. The white papers attempt to portray a comprehensive picture of the computing research field detailing potential research directions, challenges and recommendations.*

*This material is based upon work supported by the National Science Foundation under Grant No. 1734706. Any opinions, findings, and conclusions or recommendations expressed in this material are those of the authors and do not necessarily reflect the views of the National Science Foundation.*

*For citation use: Bertino E., Bliss D., Peterson L., Lopresti D., & Schulzrinne H. (2020) Computing Research Challenges in Next Generation Wireless Networking.*
*https://cra.org/ccc/resources/ccc-led-whitepapers/#2020-quadrennial-papers*